\def\year{2023}\relax
\documentclass[letterpaper]{article} 
\usepackage{aaai22}  
\usepackage{times}  
\usepackage{helvet}  
\usepackage{courier}  
\usepackage[hyphens]{url}  
\usepackage{graphicx} 
\usepackage{natbib}  
\usepackage{caption} 
\DeclareCaptionStyle{ruled}{labelfont=normalfont,labelsep=colon,strut=off} 
\usepackage{soul}
\usepackage{xcolor}

\usepackage{newfloat}
\usepackage{listings}
\usepackage[caption=false]{subfig}
\title{Tracking Fringe and Coordinated Activity on Twitter \\
Leading Up To the US Capitol Attack}
\author {
    Padinjaredath Suresh Vishnuprasad,\textsuperscript{\rm 1,\rm 2}
    Gianluca Nogara,\textsuperscript{\rm 1}
    Felipe Cardoso,\textsuperscript{\rm 1}\\
    Stefano Cresci,\textsuperscript{\rm 3}
    Silvia Giordano,\textsuperscript{\rm 1}
    Luca Luceri\textsuperscript{\rm 1,\rm 4}
}
\affiliations {
    \textsuperscript{\rm 1}SUPSI, Switzerland;
    \textsuperscript{\rm 2}IISER-Tirupati, India;
    \textsuperscript{\rm 3}IIT-CNR, Italy;
    \textsuperscript{\rm 4}USC Information Sciences Institute, USA\\
    vishnu.prasad@hdr.qut.edu.au, \{gianluca.nogara, silvia.giordano\}@supsi.ch,\\ cardosodesantana.89@gmail.com,
    stefano.cresci@iit.cnr.it, lluceri@isi.edu
}

\usepackage{bibentry}

\begin{document}

\maketitle

\begin{abstract}

The aftermath of the 2020 US Presidential Election witnessed an unprecedented attack on the democratic values of the country through the violent insurrection at Capitol Hill on January 6th, 2021. The attack was fueled by the proliferation of conspiracy theories and misleading claims about the integrity of the election pushed by political elites and fringe communities on social media. In this study, we explore the evolution of fringe content and conspiracy theories on Twitter in the seven months leading up to the Capitol attack. We examine the suspicious coordinated activity carried out by users sharing fringe content, finding evidence of common manipulation techniques ranging from targeted amplification to manufactured consensus, which went likely undetected by Twitter. Further, we 
map out the temporal evolution of, and the relationship between, fringe and conspiracy theories, which eventually coalesced into the rhetoric
of a stolen election, with the hashtag \#stopthesteal, alongside QAnon-specific narratives. Our findings highlight how social media platforms offer fertile ground for the widespread proliferation of conspiracies during major societal events, which can potentially lead to offline actions and organized violence.
\end{abstract}

\section{Introduction}
The 2020 US Presidential Election took place against the backdrop of incumbent president Donald Trump's impeachments, the COVID-19 pandemic, and racial unrest of the Black Lives Matter movement. In such a tumultuous scenario, Donald Trump challenged the integrity of the elections by diffusing claims of alleged, widespread voter fraud and claiming his victory despite the vote count \cite{universitythe}. 
Activism from fringe groups and Trump supporters gained traction online in the months following the election, calling for violence against those who opposed Trump’s claims 
\cite{pennycook_2021_research}. Following Trump's instruction (``Be there, will be wild!'') in one of his prominent tweets,\footnote{https://www.justsecurity.org/84551/important-elements-of-the-january-6th-report/} on January 6th,  2021, as Congress convened to ratify the electoral college results, his supporters attacked the US Capitol, breaching the Capitol Building and causing five deaths and multiple injuries \cite{universitythe}.

The US Capitol attack was seen as an attempted \emph{coup d'\'etat} \cite{washington_politifact}, organized and promoted online, with violent actions and repercussions offline \cite{universitythe,luceri2021social}. In the aftermath of the attack, the US House of Representatives established a Select Committee to investigate the Capitol attack, and the role of social media in relation to the “spread of misinformation, efforts to overturn the 2020 election, domestic violent extremism, and foreign influence in the 2020 election”.\footnote{https://techpolicy.press/january-6-committee-issues-subpoenas-to-social-media-platforms/} In their final report,\footnote{https://perma.cc/ZA6R-JRF2} released on December 22nd, 2022, the Select Committee examined former President Donald Trump’s conduct to invalidate the election results and foster the attack to the Capitol.\footnote{https://www.justsecurity.org/84658/insiders-view-of-the-january-6th-committees-social-media-investigation/} Besides Trump’s potentially pivotal role in driving the attack, which is out of the scope of this paper, the report highlights how Twitter, among all the subpoenaed social media platforms, served as the main channel for Trump, his supporters, and fringe groups, such as QAnon, to amplify conspiracy theories and plan the assault.
The attack ensued by large-scale deplatforming, with Twitter (until November 2022), Facebook, Instagram, and Snapchat, among the others, banning Donald Trump and many of his followers, who flocked together to alt-tech social media platforms \cite{pennycook_2021_research}. 

The research community examined the effects of these interventions \cite{horta2021platform,trujillo2022make} and studied online conversations surrounding the election from different angles.
The Election Integrity Partnership (EIP) released an observational study tracking misleading claims related to the integrity of the electoral system from the run-up to the aftermath of the election~\cite{universitythe}. 
Among all the fringe groups, the QAnon conspiracy played a relevant role in steering conversations along the line of the election fraud rhetoric \cite{luceri2021social, sharma2022characterizing}. While a large body of research focused on the voting event, a longitudinal study that examines 
the unfolding of conspiracy and fringe narratives leading up to the Capitol attack is still lacking.\footnote{For the sake of simplicity, we use the terms ``fringe narratives" and ``conspiracy theories" interchangeably throughout the paper. However, in our results, we clarify and distinguish these two distinct instances of problematic content.}

In this paper, we investigate the proliferation and coordinated amplification of conspiratorial content on Twitter in the months preceding the 2020 US Election and the Capitol attack.
Leveraging a dataset of over 500M election-related tweets gathered from July 2020 to January 2021, we deconstruct the \emph{Twittersphere} to investigate the temporal evolution of, and the relationship between, fringe narratives. 
Further, we examine the potential use of coordinated and manipulative actions to amplify content and create an illusion of consensus around conspiracies. 
In particular, we address the following research questions (RQs):
\begin{itemize}
    \item[\textbf{RQ1:}] \emph{Were fringe narratives amplified employing manipulation techniques? Who were the actors responsible for the manipulation, and which stories were pushed?}
    \item[\textbf{RQ2:}] \emph{What was the relationship between the various fringe narratives and conspiracy theories? How did they evolve over time and converge at the time of the Capitol attack?}
\end{itemize}

To address these RQs, we rely on and build upon computational models to chart the landscape of conspiratorial and coordinated activity surrounding the 2020 US Election, leading up to the January 6th attack, providing the following contributions:
\begin{itemize}
    \item Employing existing approaches for the detection of coordinated and inauthentic activity~\cite{pacheco2021uncovering,nizzoli2021coordinated,weber2021amplifying}, we found evidence of manipulation attempts to amplify and elicit an illusion of consensus around conspiratorial content. 
    \item Leveraging manually curated annotations \cite{kennedy2022repeat,voterfraud2020}, we observed how the ``Stop the Steal" rhetoric was by far the narrative mostly amplified by the coordinated activity. Further, we uncovered  and verified that the vast majority of users involved in such manipulation actions were supporting and promoting claims about voter fraud \cite{voterfraud2020}.
    \item We built on prior work \cite{mit} to map out the temporal evolution of fringe narratives and to draw a broad picture of their relationships. We discovered that conspiratorial and fringe content coalesced into two distinct, yet interlaced, meta-narratives embodying the ``stolen election" rhetoric and the QAnon conspiracy. Our results experimentally confirm the observational findings released by the \citet{universitythe}.
\end{itemize}

\section{Related Work}

\subsection{Misinformation and the 2020 US Election}
The 2020 US Presidential Election saw an unprecedented number of false claims and misinformation about voter fraud, including the self-proclamation of Donald Trump as the real winner~\cite{chen2021covid,khudabukhsh2022fringe}. 
In this context, many studies tracked the spread of electoral misinformation in the run-up to the election and its aftermath. 
Diverse collaborative and participatory schemes were used to spread questionable narratives~\cite{universitythe}. 
Some of these were \textit{bottom-up}, where influential media, political elites, and social media influencers broadcasted content produced by average users. Other times instead, the patterns of diffusion were \textit{top-down}, involving large numbers of users, automated~\cite{ferrara_2020_characterizing} or otherwise~\cite{jachim2021trollhunter2020}, that coordinated to amplify narratives proposed by influential users. 

Among the plethora of intertwined false narratives, ``Stop the Steal" gave rise to an advocacy organization responsible for multiple riots across the country \cite{luceri2021social}. The ``Stop the Steal" campaign spread across multiple platforms, including Facebook and Twitter, pushing allegations that the election was stolen and inciting Trump supporters to overturn the result~\cite{benkler2020mail,universitythe}: 
One year after the election, up to one-third of the US population still believed 
that the electoral outcome was not to be trusted \cite{matatov2022stop,pennycook_2021_research}. The large body of work on the 2020 US Election also resulted in the collection and publication of several resources, such as social media datasets of annotated posts and users involved in the spread of election misinformation~\cite{chen_2022_2020,kennedy2022repeat,voterfraud2020}. In the present study, we leverage a combination of existing resources to provide unprecedented insights into the coordinated strategies used to amplify false narratives and how these contributed to sparking the Capitol Hill attack.

\subsection{QAnon and Fringe Conspiracy Theories}
Interest in tracing the diffusion and in identifying supporters of conspiracy theories is motivated by their detrimental effect on people's political engagement and trust in authorities~\cite{green2022online}, so much so that the interplay between the spread of conspiracy theories and major political events is long studied. QAnon was the main conspiracy involved in the US 2020 electoral debate~\cite{sharma2022characterizing,wang2022identifying}. QAnon conspiracy theories are associated with a fringe political movement that centers around the belief in an anonymous user claiming to possess a high-level security clearance within the US Government \cite{papasavva2021qoincidence}. Followers of QAnon assert that a secretive group of Satanic, cannibalistic individuals engaged in global child sex trafficking conspired against Donald Trump.
The interested reader can refer to the Appendix for a detailed overview of QAnon's theories. 

\citet{sharma2022characterizing} found that around 66\% of all users that were active in the online debate interacted with QAnon content and that far-right QAnon supporters actively tried to persuade left-leaning users. Among the many narratives that originated from QAnon, the ``Hammer and Scorecard'' and ``Dominion'' theories were particularly popular during the 2020 Election. These narratives were pushed by a combination of prominent users (e.g., Donald Trump and other politicians), far-right communities (e.g., the Proud Boys), and other average users~\cite{universitythe}, including automated accounts \cite{ferrara_2020_characterizing,luceri2021down}. 

Moreover, the proponents of these extremist conspiracies also endorsed the election-fraud claims, up to the point that QAnon conspiracies coalesced in and reinforced the ``Stop the Steal'' campaign, favoring the radicalization of Trump supporters~\cite{pennycook_2021_research,luceri2021social}. This intricate interplay poses inevitable challenges towards tracing the spread of conspiracies, understanding how they unfold~\cite{papasavva2022gospel}, and what leads users to adopt the theories \cite{wang2022identifying}.



\subsection{Coordinated Online Behavior}
A frequent characteristic of mis- and disinformation campaigns is the use of \textit{coordinated activity} to boost the spread of the shared messages~\cite{nizzoli2021coordinated}. A recent body of work thus focused on detecting and characterizing coordinated online behavior as a way to surface mis- and disinformation campaigns. The majority of existing approaches detect online coordination by identifying unexpected or exceptional similarities in the action sequences of two or more users. For example, many works model common user activities (e.g., co-retweets, temporal and linguistic similarities) to build user similarity networks where coordinated groups (i.e., communities) of users are identifiable~\cite{weber2021amplifying,pacheco2021uncovering,nizzoli2021coordinated,magelinski2022synchronized}. Some methods do not only distinguish between coordinated and non-coordinated users, as in a binary classification task, but also quantify the extent of coordination thus providing more nuanced results~\cite{nizzoli2021coordinated}.

The characterization of coordinated users and communities is another thriving area of research. This task is typically aimed at distinguishing between malicious and benign forms of online coordination~\cite{vargas2020detection,hristakieva2022spread}. In fact, coordination is a necessary ingredient not only of manipulation campaigns, but also of many other social movements, including fandoms, activists, and protesters~\cite{nizzoli2021coordinated}. Longitudinal analyses of the coordinated actions of social movements revealed that both offline and online events can spark online coordination~\cite{varol2014evolution,casas2019images}. Importantly, the opposite is also true, since online coordination often precedes real-world action~\cite{steinert2015online}. These results demonstrate the importance of characterizing the multiple forms of online coordination, given the real-world consequences that they can have. To distinguish between neutral/benign and malicious online coordination, some considered the structural properties of the coordination networks~\cite{vargas2020detection,nizzoli2021coordinated,cinelli2022coordinated}, the reliability of the news they shared~\cite{cao2015organic}, or their use of propagandist language~\cite{hristakieva2022spread}, to gain insights into the objectives and intent of the coordinated users. Here, we employ state-of-the-art coordination detection frameworks~\cite{weber2021amplifying,pacheco2021uncovering,nizzoli2021coordinated} to identify users that participated in coordinated activities with the goal of spreading conspiratorial content, and we leverage curated annotations \cite{kennedy2022repeat,voterfraud2020} to characterize the messages and users involved in such manipulation attempts.


\begin{table}
\centering
\caption{Fringe hashtags analyzed in this paper.}
\label{hashtable}
\begin{tabular}{|c|c|l|} 
\hline
\textbf{US Election}   & \textbf{QAnon}               & \multicolumn{1}{c|}{\textbf{COVID-19}}     \\ 
\hline
\multicolumn{1}{|l|}{\#hammerandscorecard} & \#pizzagate         & \multicolumn{1}{c|}{\#plandemic}  \\ 
\hline
\#sharpiegate                              & \#qanon             &                                   \\ 
\hline
\#qsnatch                                  & \#qarmy             &                                   \\ 
\hline
\#stopthesteal                             & \#taketheoath       &                                   \\ 
\hline
\#dobbs                                    & \#wwg1wga           &                                   \\ 
\hline
\#dominionsoftware                         & \#projectveritas    &                                   \\ 
\hline
\#dominion                                 & \#thegreatawakening &                                   \\ 
\hline
\#hammer                                   & \#civilwar          &                                   \\ 
\hline
\#scorecard                                & \#obamagate         &                                   \\
\hline
\end{tabular}
\end{table}

\section{Dataset \& Fringe Hashtags}
\label{eda}
We leverage the first public Twitter dataset about the 2020 US Presidential Election \cite{chen_2022_2020}, which was collected using election-specific keywords and gathering messages from a set of political figures, including the running candidates. This dataset was gathered in real-time by leveraging the 1\% stream of tweets provided by Twitter’s streaming API. From the list of released tweet IDs \cite{chen_2022_2020}, we applied a rehydration process to collect the full tweet objects.
For our analyses, we use a subset of the whole dataset considering only tweets from July 2020 to January 2021, to cover an observation period that goes from the final campaign to the Capitol attack.
This results in a longitudinal dataset that contains about 568 million tweets, consisting of 52\% of retweets, 19\% of replies, 16\% of original tweets, and 13\% of quotes. The rehydration process allowed us to retrieve about 51.72\% of the tweets collected by \cite{chen_2022_2020} in our window of observation. The missing tweets were likely deleted or removed by Twitter's intervention against QAnon \cite{wang2022identifying}.

\begin{figure}[t]
    \centering
    \includegraphics[trim={0 1cm -2cm 0}, width=0.9\linewidth]{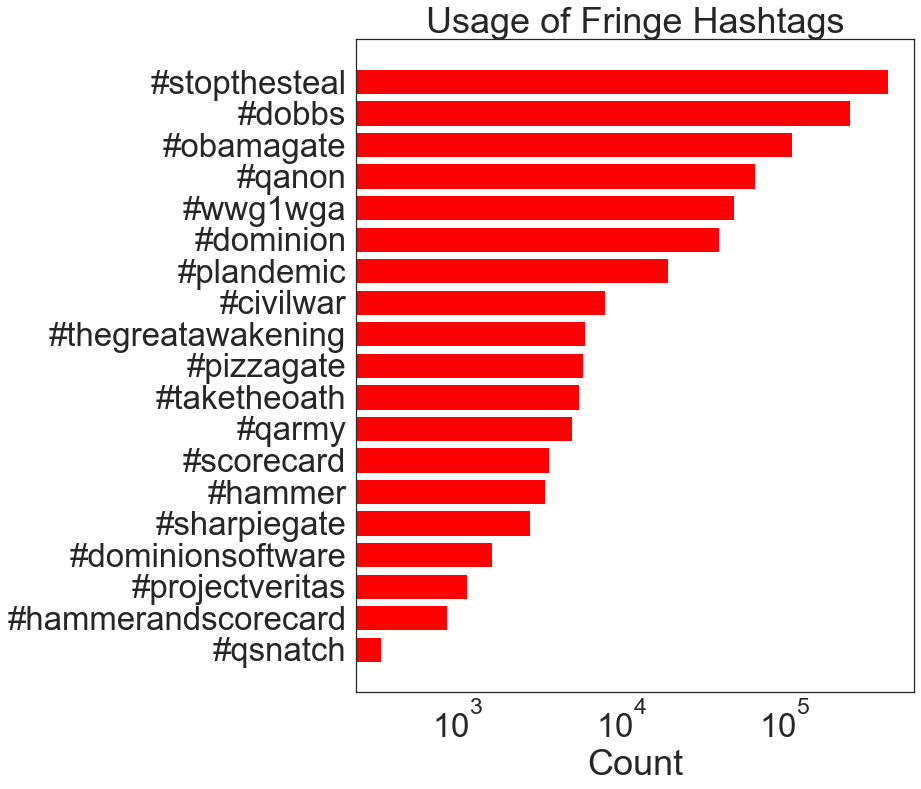}
    \caption{Distribution of fringe hashtags.}
    \label{bars}
\end{figure}

Leveraging previous analyses on the same dataset \cite{ferrara_2020_characterizing, sharma2022characterizing, wang2022identifying,chen_2022_2020} and diverse reports on the Capitol attack \cite{pennycook_2021_research, universitythe,luceri2021social}, we identified nineteen hashtags pushed by conspiracies and fringe communities, which we refer to as \emph{fringe hashtags}. In this body of literature, the fringe hashtags are commonly grouped in three broad categories related to \textit{(i)} QAnon -- hashtags originated from and propagated by the QAnon conspiracy; \textit{(ii)} US Election  -- hashtags encompassing a diverse set of conspiracies supporting the election fraud rhetoric; and \textit{(iii)} COVID-19 -- a single hashtag related to a conspiracy theory about the pandemic. These categories and corresponding hashtags are summarized in the Appendix. The three groups of fringe hashtags are displayed in Table \ref{hashtable} while Figure \ref{bars} depicts their distribution in our dataset. Notably, \#stopthesteal -- promoting the stolen election narrative, and \#dobbs -- supporting the former Fox News political commentator Lou Dobbs, were by far the most shared hashtags, followed by \#obamagate, \#qanon, \#wwg1wga, \#dominion, \#plandemic, and \#civilwar.

\section{Methodology}

To address the proposed RQs, we leverage three methodologies. We explore suspicious coordinated activity relying on existing techniques that \textit{(i)} identify networks of \emph{rapid retweet} interactions -- generally employed to amplify content generated by targeted users \cite{mazza2019rtbust,pacheco2021uncovering}, and \textit{(ii)} discover clusters of users sharing similar tweets -- usually shared with small variations and nuances to create an illusion of consensus \cite{weber2021amplifying}. We refer to these two network-based models as \emph{rapid retweet network} and \emph{copypasta network}. In regard to the suspicious coordinated activities surfaced by our methodologies, we remark that \citet{nizzoli2021coordinated} highlighted the difference between coordination and inauthenticity, admitting the existence of a spectrum of behaviors that ranges from benign to neutral and malicious coordinated activities. However, since the hashtags analyzed in our work endorse empirically disproved conspiracy theories, we posit that the coordination found with our methodologies relates to malicious and inauthentic activities carried out with the intent to misinform. 

Further, we characterize \textit{(i)} 
 the users involved in such manipulation attempts by leveraging the annotations produced by \citet{voterfraud2020}, who classified users as either promoters or detractors of voter fraud narratives; and \textit{(ii)} the tweets shared by these coordinated accounts relying on the curated dataset from \citet{kennedy2022repeat} to label messages sowing doubts in the legitimacy of the election.
 Finally, we build on the approach proposed by \citet{mit} to chart the landscape of fringe narratives by mapping out their relationship and evolution over the months leading up to the Capitol attack. We refer to this augmented model as \emph{Hashtag Temporal Evolution Mapping (HTEMap)}. 

\subsection{Rapid Retweet Network}
\label{rrtmeth}
Retweets, excluding quoted tweets, represent the quickest and easiest form of social endorsement on Twitter \cite{metaxas2015retweets}, and they signal a high level of trust and agreement with the shared message.  
Users can re-share others' content in a simple, \emph{one-click} operation, which can even be automated \cite{luceri2019evolution}. However, the simplicity behind this action led to its nefarious use in orchestrated manipulation campaigns \cite{bessi2016social}, where retweets were collaboratively employed to spread and amplify (mis-)information shared by influential users \cite{mazza2019rtbust}. We remark that quoted tweets, which are possibly used to express criticism rather than endorsement, are not considered in this analysis. 

Given the widespread misuse of retweets, in this work, we aim to identify networks of accounts that suspiciously employ retweets in a fast, repeated, and coordinated fashion.
Following the methodology presented by \citet{pacheco_2020_unveiling}, we define a rapid retweet as a re-share of a tweet that occurs in a narrow time window. Here we consider different sizes of the time window to assess the strength of this coordinated activity. According to this definition, we build a weighted \emph{rapid retweet network}, where the user that re-shared a tweet (\emph{source}) is connected to the user that generated the original content (\emph{target}) if the retweet occurred within the designated time window size, and the weight represents the number of rapid retweets between the source and target users. 
Similarly to \citet{pacheco_2020_unveiling}, we also filter out occasional instances of rapid retweets that might happen by chance by considering only links with a weight larger than one. 


\subsection{CopyPasta Network}
\label{cpmeth}
Similarly to retweets, content with high similarity can be used to resonate messages and elicit the idea of widespread agreement around an idea, a piece of information, or a misleading claim.
For the sake of simplicity, and leveraging the Internet jargon, we refer to tweets with high text similarity as \emph{copypasta tweets}. To identify such tweets, we carry out a pair-wise comparison of the text of original content shared in our dataset, thus including replies, quotes, and original tweets. We do not consider retweets in this subset as, by definition, they have exactly the same text and cannot be considered as an instance of \emph{copypasta tweet}.

Calculating pairwise similarity between a huge volume of tweets is, however, a significant resource- and time-consuming task. Similarly to \citet{pacheco_2020_unveiling}, we sort tweets in chronological order and compare only messages shared in close time proximity, considering a sliding window of ten tweets. For each time window, we construct a 10$\times$10 matrix of tweet similarity scores, which are in turn computed using the Ratclif/Obershelp algorithm \cite{ratclif_1988_pattern}. 
This computation forms the basis to build the \emph{copypasta network}, defined as an undirected network of tweets linked to each other if their similarity score is above a certain threshold. 
As we thoroughly show in the following sections, this threshold is not chosen arbitrarily, but rather it is adequately selected by observing the 
distribution of similarity scores of both fringe and generic tweets.

\subsection{Characterizing Users and Tweets} 

\begin{figure}[t!]
    \center
    \includegraphics[trim={0 0 2cm 0}, width=.9\linewidth]{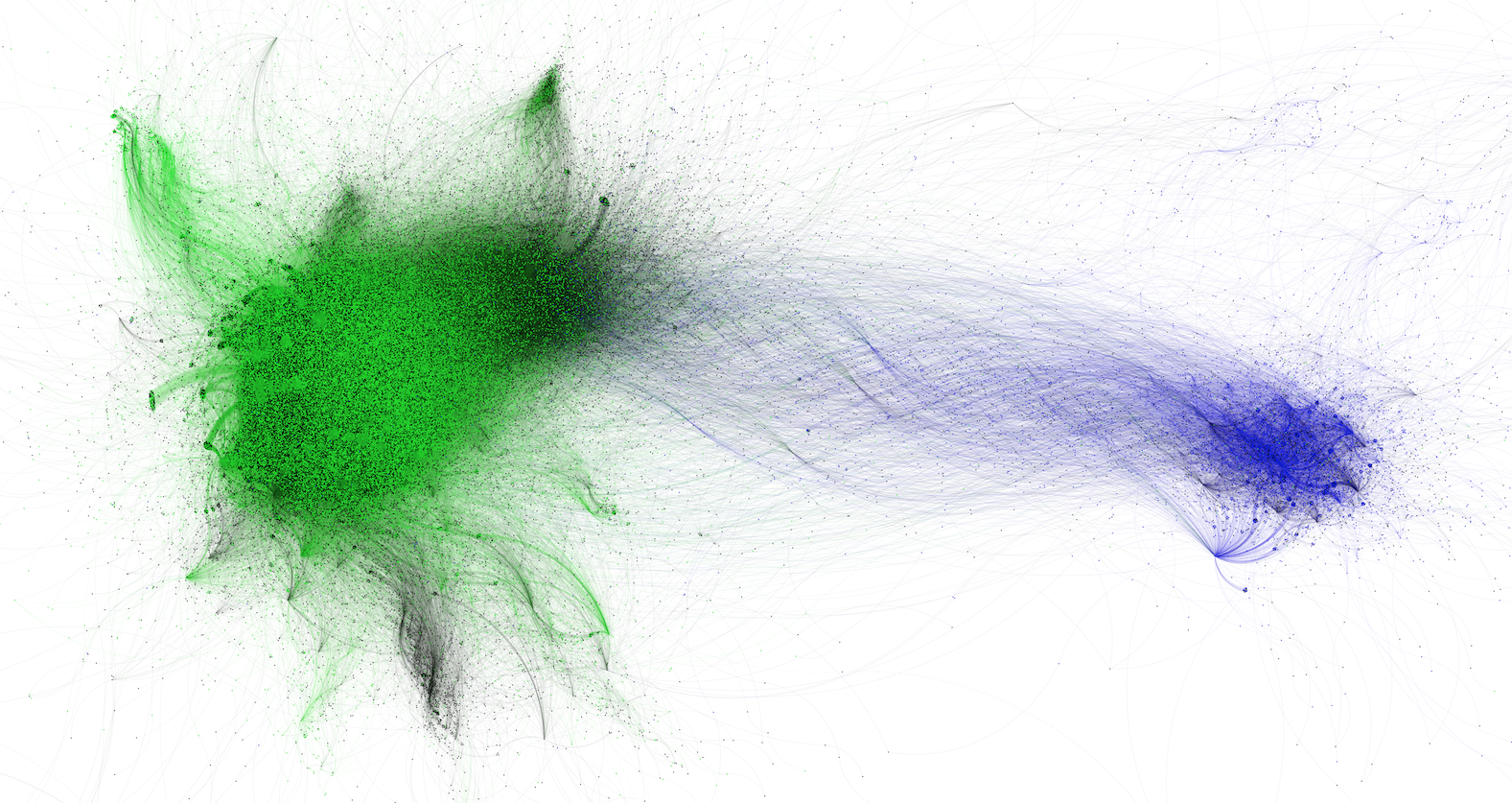}
    \caption{Retweet network of detractors (blue) and promoters (green) of the voter fraud conspiracy, along with unlabeled users (black). The node size represents the in-degree, i.e., the number of times the user was retweeted.}
    \label{prodet}

\end{figure}

To characterize the users involved in the spread of fringe narratives, we leverage a publicly-available annotation released by \citet{voterfraud2020}. There, the authors classify users as either \emph{detractors} or \emph{promoters} of the voter fraud rhetoric and specific conspiracy theories.
Leveraging a rich set of information (shared URLs, images, and YouTube videos), they discovered five distinct communities, where \emph{detractors} sit in a unique community and \emph{promoter} populate the other four communities. Here we use the annotations of \citet{voterfraud2020} to label users as \emph{detractors} or \emph{promoters} and to understand their role in the coordinated inauthentic activities described before. The overlap of users with \citet{voterfraud2020} resulted in a labeling of 63.25\% of the users in our dataset and as we could expect, the majority of those (81.5\%) represent \emph{promoters} of the voter fraud rhetoric. It is worth noting the relevant fraction of users (18.5\%) labeled as \emph{detractors} in our dataset. This is not surprising, however, given that \emph{detractors} of conspiracies also make use of controversial hashtags and quote misleading claims to attack or convert fringe users \cite{wang2022identifying}. \citet{voterfraud2020} also established a connection between users' classification and their political alignment, with AllSides and Media Bias/FactCheck serving as references. Their research revealed that \emph{promoters} generally align with a conservative orientation, whereas \emph{detractors} align with a liberal orientation.

Figure \ref{prodet} represents the interactions between detractors and promoters with a directed, weighted retweet network, where the nodes are users, edges represent retweets, and the weight is proportional to the frequency of retweets between each pair of nodes. The network in Figure \ref{prodet} is visualized using a force-directed layout (ForceAtlas2), where nodes repulse each other, whereas edges attract their nodes. 
This results in a natural split of the network into two communities entirely populated by users with the same opinion about the voter fraud rhetoric. 
Unlabeled users are balanced between the community of promoters (53.23\%) and the community of detractors (45.17\%).\footnote{Note that the percentages do not sum up to 100\% as a small fraction of unlabeled nodes does not belong to any community.} 
In the next sections, we delineate the role of these two distinct communities in the manipulation attempts of amplifying fringe narratives.

Additionally, we characterize a subset of tweets in our dataset by leveraging the annotations curated by \citet{kennedy2022repeat}, which manually associated tweets to stories on the delegitimization of the election. Since the analysis by \citet{kennedy2022repeat} involves a much shorter time window than that analyzed by us
(i.e., it overlaps with ours for 90 days, which corresponds to 28\% of the whole time covered by our data), their annotations only cover 21.48\% of all tweets in our dataset. Nonetheless, when combined with our analyses of the manipulation tactics adopted in the run-up to the election, such annotations allow us to understand which stories were artificially amplified. It is important to note that we use the dataset provided by \cite{kennedy2022repeat} to complement our study based on hashtags with a meta-level analysis of manually-annotated tweets. 
Notably, the annotations by \citet{kennedy2022repeat} are also characterized by a large agreement between the human coders, as testified by a Fleiss’s $\kappa$ score = 0.64. The top five misleading stories in our dataset are related to the ``Stop the Steal" narrative, named \emph{Stop the Steal Pushed} (46.71\%) and \emph{Stop the Steal rallies} (4.03\%) by \citet{kennedy2022repeat}, followed by the \emph{Dominion} (38.87\%), \emph{Hammer and Scorecard} (4.52\%), and \emph{Sharpiegate} (1.54\%) conspiracy theories.

\subsection{HTEMap Model}
\label{htemap}

\begin{figure}[t!]
    \centering
    \includegraphics[trim={0 1cm 0 0}, width=0.5\textwidth]{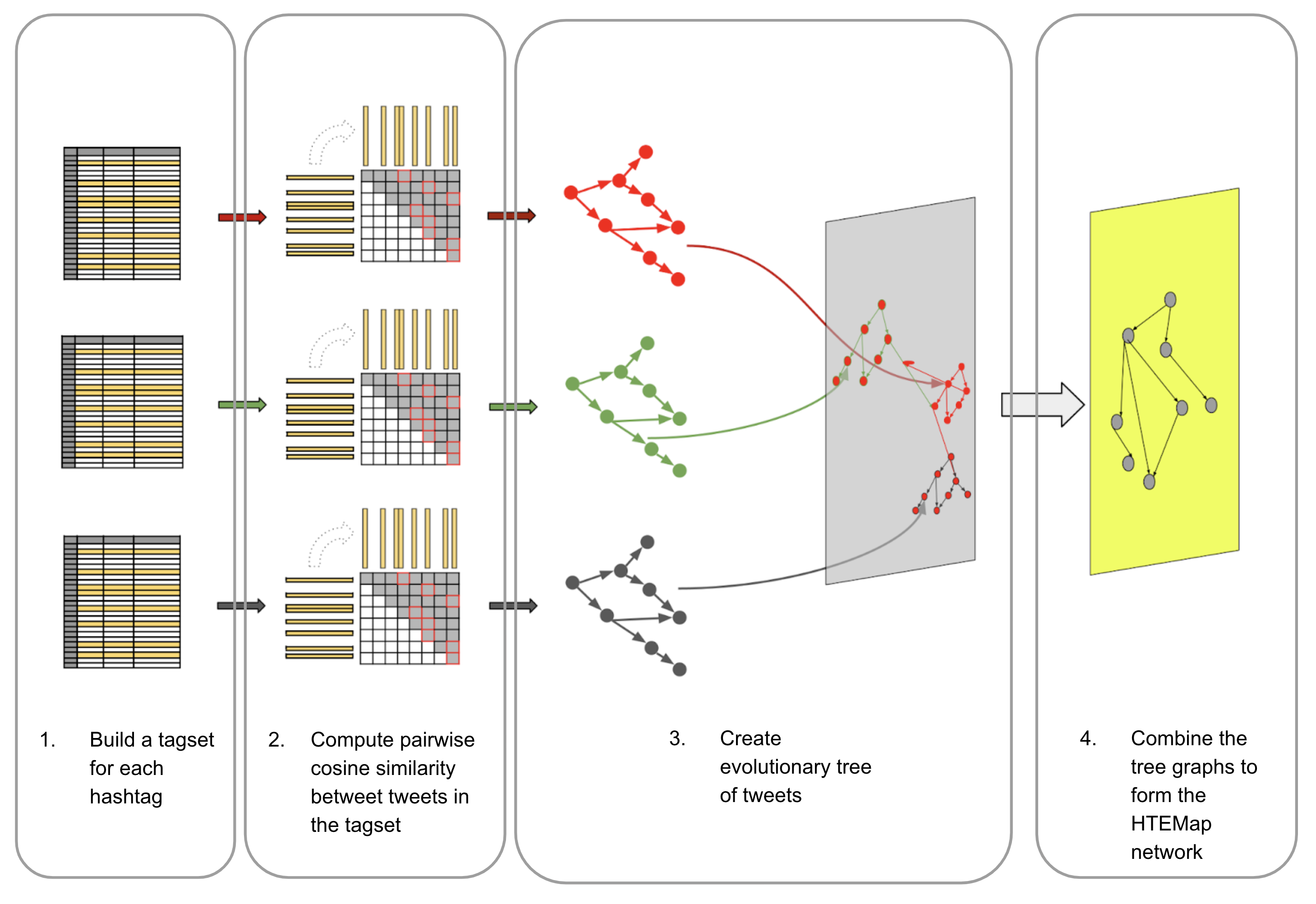}
    \caption[HTEMap Model]{The HTEMap Model.}
    \label{mth}
\end{figure}

To delineate the relationships between the various fringe narratives
and map out their temporal evolution, we augment the model proposed in \citet{mit} to introduce the \emph{Hashtag Temporal Evolution Mapping (HTEMap)} model and construct a temporal network of hashtags. In particular, HTEMap infers an evolutionary trajectory of hashtags by first charting the relationship between tweets and hashtags across time as in \citet{mit}, and then building a co-occurrence network of hashtags. The architecture of the model is displayed in Figure \ref{mth} and consists of four steps. 

First, for every hashtag, we extract a hashtag dataset (in short \emph{tagset}), which includes the tweets embedding the target hashtag, sorted in reverse-chronological order. 

Second, we perform a pairwise comparison of the tweets in the tagset. We represent each tweet with a $n$-dimensional vector (in our setting $n=19$, given that we are considering 19 fringe hashtags) whose elements, $b_k$, represent the number of occurrences of the $k$-th hashtag in the tweet. 
We build a similarity matrix $S$, where every element $S_{i,j}$ represents
the cosine similarity between the $n$-dimensional representations of tweets $(i)$ and $(j)$.  



\begin{figure*}[t!]
    \center
    \includegraphics[width=\textwidth]{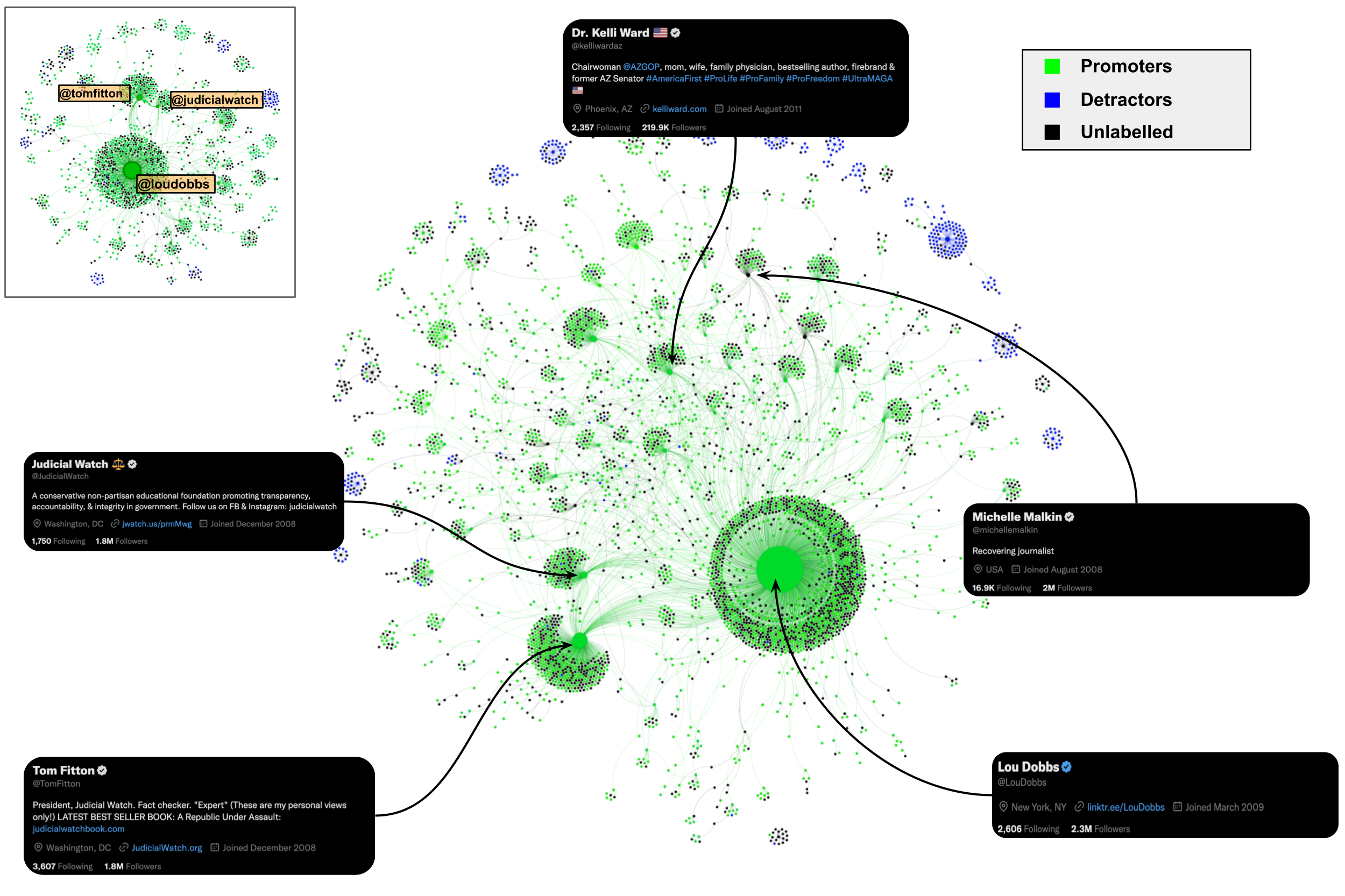}
    \caption{The rapid retweet network. Node size represents the in-degree, whereas the node color characterizes users as promoters or detractors \cite{voterfraud2020}. The inset represents an instance of the rapid retweet network with a shorter time window (30sec).}
\label{rrt}
\end{figure*}


Third, we use the similarity matrix $S$ to construct a graph of tweets. Specifically, we consider only the upper-triangular matrix of $S$ to select the most similar pair of tweets and connect their corresponding nodes with a directed edge that goes from the older one to the newer one. 
Once performed this operation for every tweet in the tagset, we obtain an evolutionary tree for the target hashtag. By repeating the previous steps for all the hashtags of interest, we obtain a rich suite of distinct, temporally directed \emph{tree graphs} that connects tweets containing the hashtags under analysis. It should be noted that, as tweets can have multiple hashtags, a tweet might be included in multiple tree graphs.

Finally, to infer the relationship between the hashtags, we construct a bipartite network composed of tweets and hashtags. We combine the nineteen tree graphs to get the \emph{tweet layer} of the bipartite network, where every node represents a unique tweet and edges connect tweets as in the tree graphs. The \emph{hashtag layer} comprises the examined nineteen hashtags. In the bipartite network, an edge exists between a tweet and a hashtag if the former embeds the latter. 
Thereby, the topology of the diverse tree graphs modulates the connections (and their weights) in the bipartite network. From this bipartite graph, we extract a hashtag network, named \emph{HTEMap network}, based on the co-occurrence of the hashtags in the tweet layer, which encompasses both the temporal evolution and relationship between the fringe hashtags.

\section{Results}

\begin{figure*}[t]
    \center
    \includegraphics[trim={0 0 0 0}, width=0.9\textwidth]{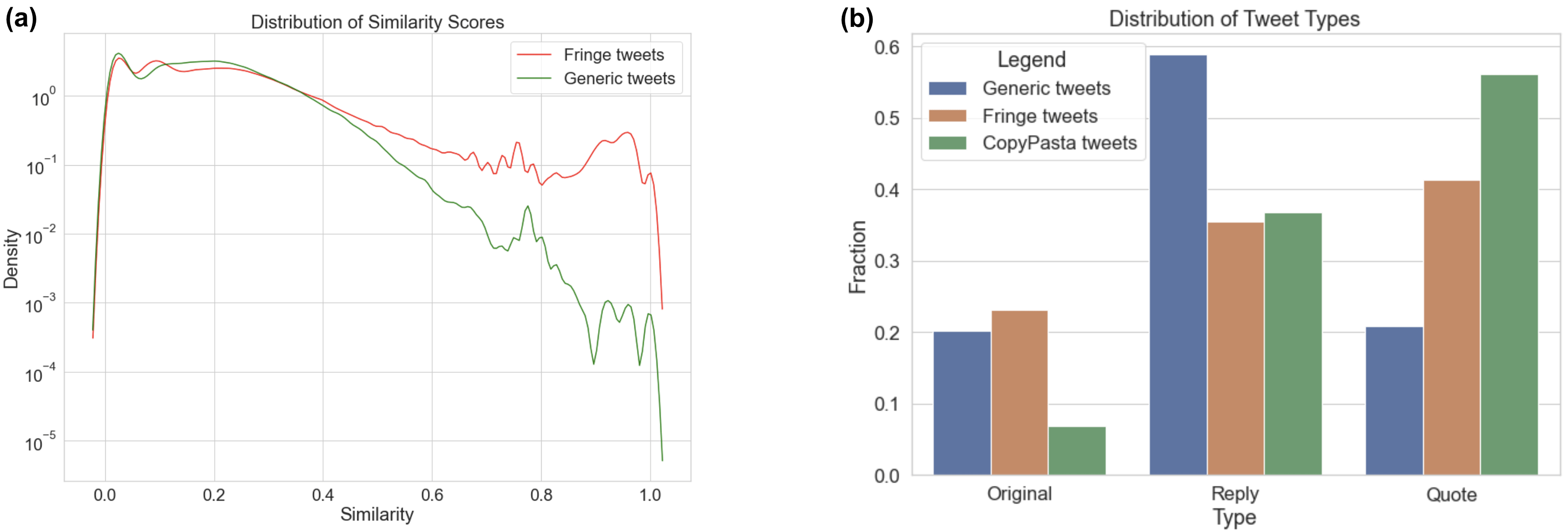}
    \caption{CopyPasta Activity: (a) Distributions of similarity scores of generic tweets and fringe tweets. \linebreak (b) Distribution of sharing activities across generic, fringe, and copypasta tweets.}
    \label{cptype}
\end{figure*}

We first discuss results related to the inauthentic activities carried out by coordinated accounts to push conspiratorial content, and then we portray the broad picture of the relationship and temporal evolution of fringe hashtags generated by the HTEMap model.

\label{results}
\subsection{Coordinated Activity (RQ1)}

To address RQ1, we analyze the techniques used to amplify fringe narratives and we characterize the tweets and users involved in coordinated inauthentic activities.

\textbf{Rapid Retweet Network.} We commence our analysis by exploring the \emph{rapid retweet network}, which offers an \emph{interactome} of the interactions between the users that coordinately amplified fringe content by means of rapid retweets. As a starting point, we consider a time window of 60 seconds to define a rapid retweet.
Figure \ref{rrt} portrays the force-directed layout of the rapid retweet network, where nodes represent users and edges indicate rapid retweets. 
The node size is proportional to the in-degree (i.e., the number of times a user was retweeted), whereas nodes are colored according to the annotations from \citet{voterfraud2020}. 
The thickness of the edges is proportional to the number of rapid retweets between nodes, while the color abides by the color of the source node (i.e., the author of the retweet). Figure \ref{rrt} highlights that the rapid retweet network is composed of network structures exhibiting a star topology. In our scenario, this means that a single user is retweeted by a multitude of other accounts. Lowering the size of the time window to 30 seconds, from the inset in Figure \ref{rrt}, we also notice how the skeletal structure of the rapid retweet network persists, with some of the most prominent accounts still at the center of their star topology and retweeted by a non-negligible number of followers, suggesting a robust, coordinated effort to target content generated by these accounts. Notably, star-like interaction structures were found in recent literature as evidence left behind by coordinated online manipulations~\cite{nizzoli2020charting}. By manually inspecting the users sitting in the center of such amplification bubbles, we identify some prominent and influential actors, including verified accounts with millions of followers. Among the others, we recognize various Republican politicians (e.g., Dr. Kelli Ward, and Ed Martin), alt-right media outlets (e.g., RSB Network), and political commentators such as Tom Fitton -- the president of Judicial Watch, a right-leaning activist group -- and the former Fox News host, Lou Dobbs -- also known for his controversial opinion about COVID-19 vaccines and his support to the voter fraud conspiracy theory \cite{Battaglio_2021}. In this regard, it is also worth noting that \#dobbs is one of the most prevalent hashtags in our dataset, as shown in Figure \ref{bars}.  

Further, looking at the node color, based on the annotations from \cite{voterfraud2020}, we find that 83\% of the users involved in the rapid retweet network are labeled as promoters of the voter fraud conspiracy theory, which highlights the propensity of these accounts in trusting conspiratorial content and their alleged intent of resonating such narratives in a coordinated fashion. Besides characterizing the actors responsible for this targeted amplification, we are interested in uncovering the narratives pushed through their activity. The hashtag \#stopthesteal (43.08\%) results to be the most shared, followed by \#dobbs (31.4\%), \#obamagate (14.12\%), and \#qanon (3.6\%). Leveraging the annotations from \citet{kennedy2022repeat}, we identified only a small set of misleading stories in the tweets boosted by this coordinated action, with the ``Stop the Steal" rhetoric standing out with 15,285 tweets, followed by the \emph{Dominion} (8,663 tweets), and \emph{Hammer and Scorecard} (1,111 tweets) conspiracies.

\begin{figure*}[ht]
    \center
    \includegraphics[width=\textwidth]{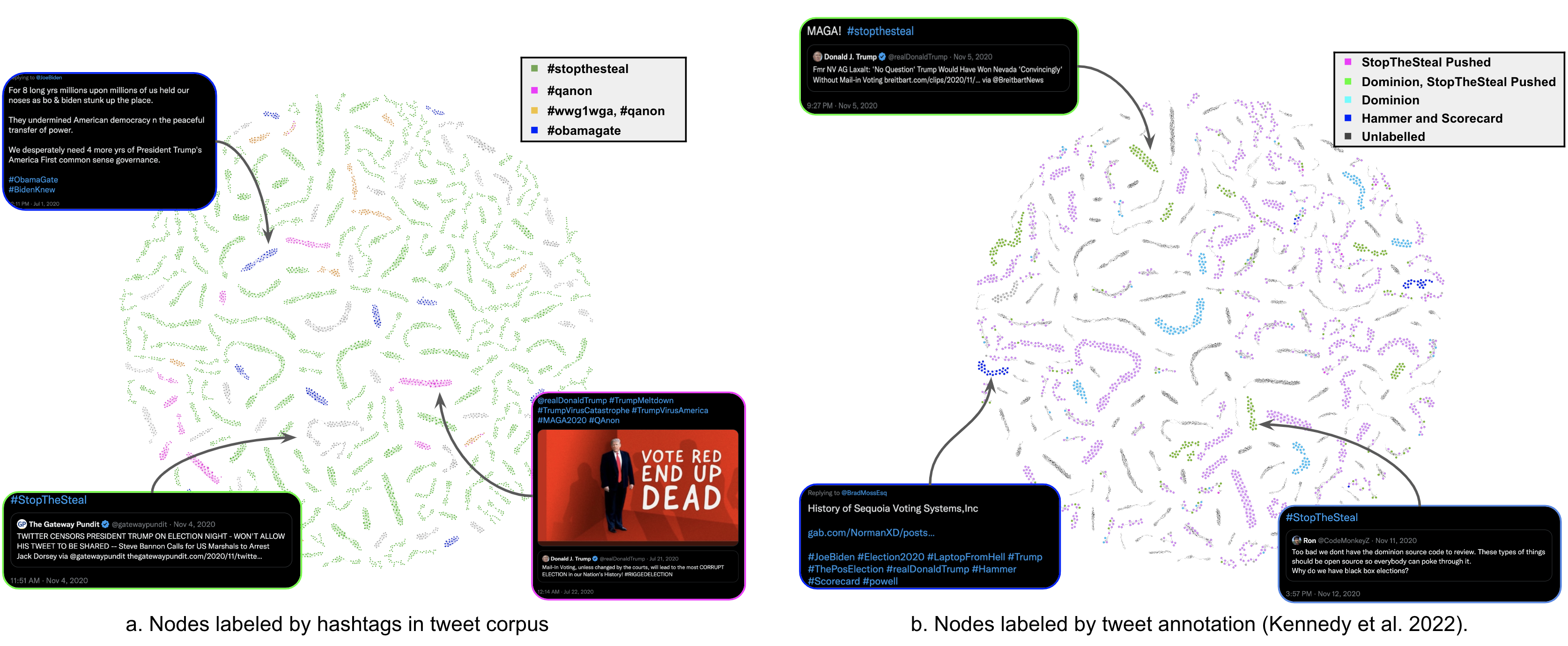}
    \caption{Copypasta network with tweets labeled by (a) hashtags, and (b) misleading claims about the election \cite{kennedy2022repeat}. A qualitative analysis of the shared content reveals that these tweets promote calls to action, support, and petitions.}
    \label{cpt}

\end{figure*}

\textbf{Copypasta Network.} To build the copypasta network, we first need to select an adequate threshold to identify similar tweets, which we also name \emph{copypasta tweets}. To this aim, we observe the distribution of similarity scores in the analyzed tweets.
From Figure \ref{cptype}(a), we note a bimodal distribution in the similarity scores of both fringe tweets (i.e., messages containing fringe hashtags) and a randomly sampled subset of messages (from our rehydrated dataset) that do not contain fringe hashtags. This surfaces a natural threshold $=0.7$ for identifying copypasta tweets, which is also consistent with previous work \cite{pacheco_2020_unveiling, topinka2022politics}. 
Interestingly, the distribution related to the sample of generic tweets shows less prevalence of copypasta tweets, indicating that this manipulation technique was mainly used to push fringe narratives.



In Figure \ref{cptype}(b), we further characterize the tweets containing fringe hashtags in contrast with generic tweets, also considering the copypasta activity. In particular, we disentangle tweets in different sharing activities (original tweets, replies, quotes) and measure their proportion for three different sets of tweets: generic tweets, fringe tweets, and fringe tweets with copypasta activity (from now on, simply \emph{copypasta tweets}). We do not consider retweets in this evaluation to carry out a fair comparison with copypasta tweets, which cannot include retweets by definition. 

From Figure \ref{cptype}(b), we observe remarkably high proportions of quotes in the copypasta activity, suggesting that this sharing activity might be used to diffuse problematic content, as further indicated by the suspiciously high volume of fringe tweets.
Both fringe and copypasta tweets include a significant fraction of replies (despite being less prominent with respect to generic tweets), which resembles the activity of misbehaving users during recent geo-political events \cite{pierri2022does}.
Interestingly, the volume of original content shared in copypasta tweets is narrow, which might suggest the potential employment of automated accounts to reproduce existing content and broadcast it via replies and quotes. This investigation is, however, out of the scope of this paper.




Based on these insights, we construct a copypasta network incorporating tweets with similarity $> 0.7$. Figure \ref{cpt}(a) shows the copypasta network, where nodes represent tweets, which are connected if they have a similarity score $> 0.7$. We observe clusters of tweets independent of each other, representing chains of copypasta activity.
Each cluster represents a set of tweets sharing the same or very similar text. Nodes are colored based on the hashtag (or combination of hashtags) embedded in the tweet. We note a huge volume of copypasta tweets embedding the \#stopthesteal hashtag (76.25\%), indicating how this hashtag was heavily amplified by means of different manipulation techniques. 

To have a better understanding of the narratives amplified by this coordinated activity, we leverage the annotations curated by \citet{kennedy2022repeat}. In Figure \ref{cpt}(b), we highlight the tweets that pushed claims about election delegitimization by coloring network nodes based on the specific theory shared.
It can be noticed how a relevant fraction ($\sim$60\%) of clusters contains at least one colored node, which suggests a relevant prevalence of misleading claims shared in a coordinated way through copypasta activity. Similarly to the results related to the rapid retweet network, ``Stop the Steal" represents the narrative most amplified by coordinated accounts (83\% of the tweets including misleading claims). 

By manually inspecting the largest clusters of tweets, we observe a variety of 
conspiratorial flavors in copypasta tweets, including calls to action, petitions, support via retweets, and trolling behavior via replies.
A qualitative analysis was conducted by examining the text, embedded media objects, and URLs within the copypasta clusters. This analysis revealed that these tweets encourage various types of engagement, including calls to action such as marches, tweet-storms, and link-sharing. They also provide support through donations, in-person rallying, and organizing activities, as well as petitions involving legal challenges, fundraising, and engagement with elected Republican representatives. Figure \ref{cpt} showcases several examples of these tweets, with usernames hidden to protect privacy.

Focusing on the actors responsible for this coordinated activity, we find that 95\% of the users involved in the spread of copypasta tweets are labeled as promoters of the voter fraud conspiracy theory \cite{voterfraud2020}, once again highlighting their pivotal role in this manipulation. Interestingly, this set of users has a tiny overlap with the accounts involved in the rapid retweet network, with only 48 accounts engaged in both coordinated activities, indicating that diverse groups of users exhibit different behaviors. This finding aligns with a similar pattern discovered in \cite{wang2022identifying} regarding users engaging with the QAnon conspiracy.


 
\subsection{Temporal Evolution (RQ2)}
\label{resrq2}

\begin{figure}[t!]
    \center
    \includegraphics[width=1\linewidth]{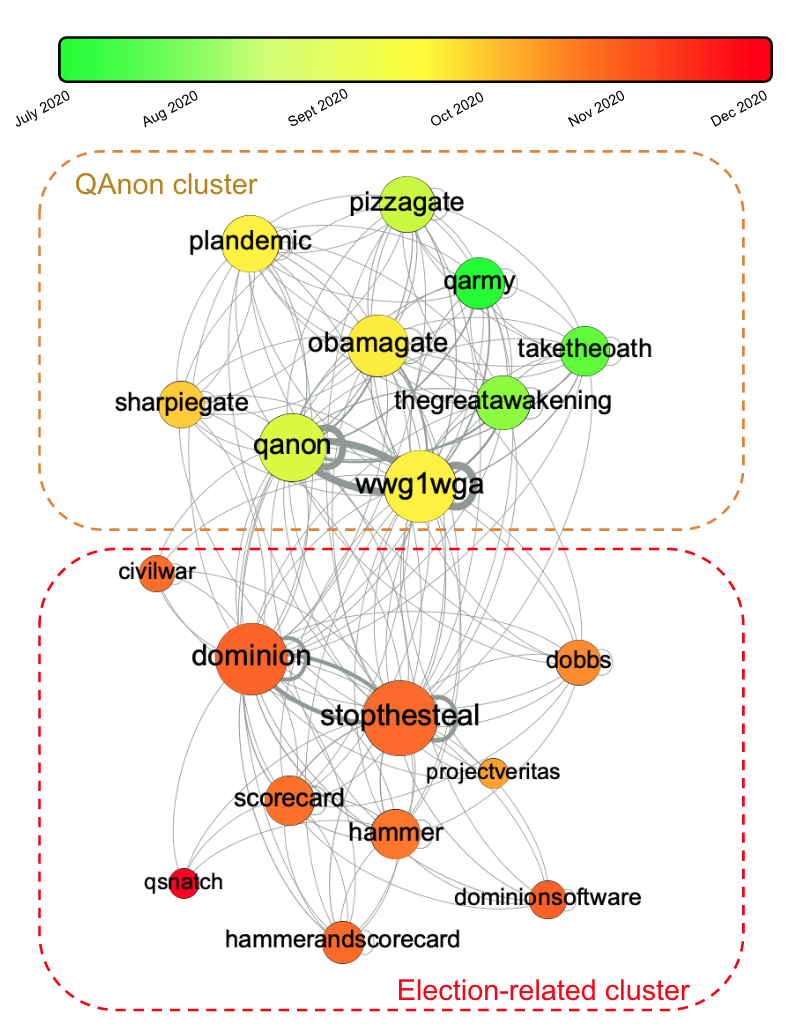}
    \caption{The HTEMap network.}
    \label{l2b}
\end{figure}

\begin{figure}[ht]
    \center
    \includegraphics[width=0.5\textwidth]{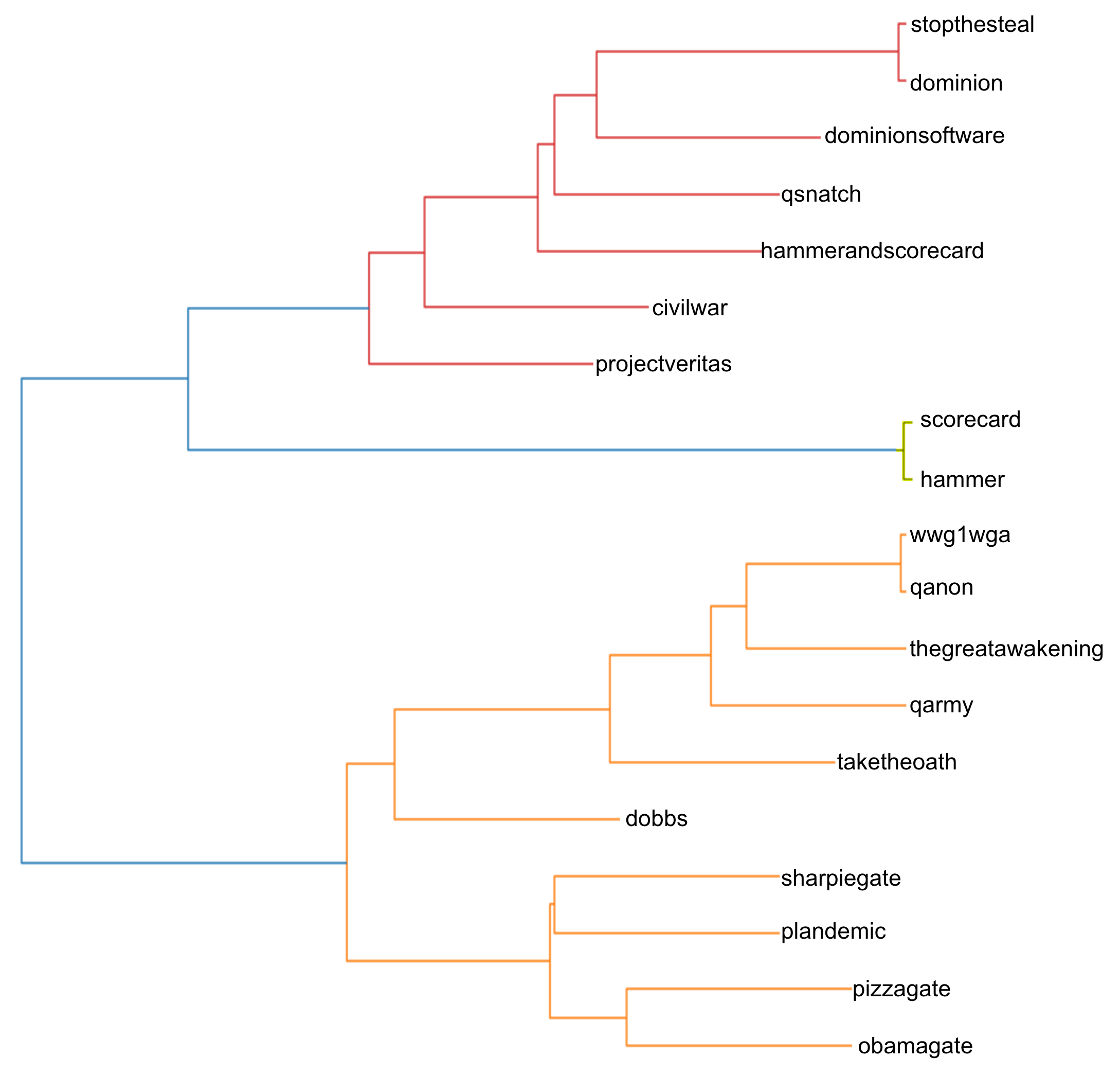}
    \caption{Evolutionary dendrogram of the fringe hashtags.}
    \label{l2}
\end{figure}

To answer RQ2, we leverage the HTEMap model for charting the landscape of the fringe hashtags, their relationship, and temporal evolution across the seven months preceding the Capitol attack.
The HTEMap model returns a temporal network of hashtags, namely the HTEMap network, which is displayed in Figure \ref{l2b} with a force-directed layout. The node size is proportional to the eigenvector centrality, and the edge weight represents the frequency of co-occurrence between the two hashtags in the HTEMap network. Similarly to what was done in recent longitudinal analyses of information operations~\cite{kriel2019reverse}, the temporal dimension is represented by the node color. It should be noted that the HTEMap network encompasses by design the temporal aspect in the construction of the evolutionary trees. Here, for visualization purposes, we consider the median of the tweets' creation date for every set of chronologically-sorted tweets containing the hashtags under analysis.  

From Figure \ref{l2b}, we observe two distinct communities, also confirmed through community extraction performed with the Louvain methodology \cite{blondel_2008_fast}, suggesting that there are two distinct, yet interconnected narratives shared in the \emph{Twittersphere} during our period of observation. Interestingly,
each community almost entirely corresponds to one of the categories of hashtags detailed in Table \ref{hashtable}, reflecting election-specific conspiracies and QAnon narratives. The \#plandemic hashtag sits in the QAnon cluster, allegedly because of the connection with other QAnon conspiracies. 
We also recognize that the most central nodes (in terms of eigenvector centrality) are \#qanon, \#wwg1wga, \#stopthesteal, and \#dominion, which connect Election-specific theories with the QAnon conspiracy.
It is also worth noting the role of \#dobbs and \#civilwar in bridging the two groups of conspiracies together. The former was amplified with the corresponding account (@LouDobbs) with a targeted amplification activity, as shown in the results related to the rapid retweet network. Instead, the latter was used by several influential spreaders to advertise the January attack and, as confirmed by our results, became viral on Twitter during the first days of November \cite{universitythe,luceri2021social}.
The interconnection of \#civilwar with the QAnon theories further confirms the involvement of the conspiracy in the organized violence behind the Capitol attack.\footnote{https://www.theguardian.com/us-news/2021/nov/17/qanon-shaman-jacob-chansley-sentenced-capitol-attack-role}
Taking the temporal dimension into account, in Figure \ref{l2b} we also observe that QAnon-specific hashtags were prevalent much earlier than the election-specific ones. In fact, almost all the election-specific fringe hashtags got traction around Election day (November 3rd, 2020), which is in line with previous research \cite{universitythe}. 

To deepen our understanding of the relationships between these hashtags, we perform a hierarchical clustering based on the adjacency matrix of the HTEMap network. This results in the dendrogram of fringe narratives shown in Figure \ref{l2}. 
To interpret the dendrogram, we focus on the width at which any two hashtags are branched together. We observe that the branch that joins \#stopthesteal and \#dominion is one of the smallest in the dendrogram, suggesting that they are among the most similar nodes in the HTEMap network. We recall that similarity in this context captures different dimensions embedded in the HTEMap network -- that is, the temporal evolution and the topological co-occurrence of the hashtags. 

From Figure \ref{l2}, we observe the emergence of two distinct branches of conspiracy theories. 
This further confirms a separation between the co-occurrence of election misinformation and fringe conspiracy theories.
In addition, it also surfaces the existence of two unique, yet interconnected, narratives in the tweetspace. 
On the one hand, \#stopthesteal and \#dominion branch coalesce together all the election-specific conspiracies, including the 2017 QAnon theory called “Hammer and Scorecard”, which gained traction again in the weeks after the 2020 election \cite{universitythe}. On the other hand, QAnon-specific hashtags pivoted around the hashtags \#qanon and \#wwg1wga but also included the ``gate" and Covid-19 conspiracies.



\section{Discussion}

\subsection{Conclusions and Future Work}
We carried out a longitudinal analysis of the interplay between conspiracy theories and fringe narratives in the run-up to the US 2020 Presidential Election. 
By employing an innovative combination of methods, we surfaced and analyzed two widely used manipulation techniques: rapid retweets and copypasta activity. Furthermore, we leveraged existing annotated resources to provide a characterization of the users involved in such manipulations, recognizing different groups of promoters of the voter fraud rhetoric acting in diverse deceptive efforts. Finally, we reconstructed the temporal trajectory of the fringe theories that eventually coalesced into the ``Stop the Steal" narrative, pinpointing its interconnection with QAnon and the Capitol attack, also advertised with the hashtag \#civilwar. 

Among the key findings of our work is that \textit{(i)} the rapid retweet networks artificially and coordinatedly amplified multiple fringe narratives (RQ1), \textit{(ii)} the most artificially amplified users were right-wing commentators and politicians that endorsed the election fraud and QAnon conspiracies (RQ1), \textit{(iii)} more than 76\% of the copypasta tweets and more than 96\% of the users involved in copypasta, promoted the ``Stop the Steal" rhetoric (RQ2), and \textit{(iv)} the QAnon conspiracies gained traction well before the election and quickly evolved into election fraud conspiracies after the polling (RQ2). Overall, our results provide insightful answers to our research questions about the adoption of manipulation techniques to amplify fringe narratives and about the interplay between the various fringe narratives and conspiracy theories. Furthermore, they shed new light into the complex socio-technical phenomena that led to one of the most destabilizing attacks on US democracy.

In the future, we aim to extend our comprehensive analyses by also considering additional platforms and other forms of online manipulation. To this end, the possibility of complementing the temporal aspects of this work with multi-platform analyses represents a promising direction of research that holds the potential for shedding further light on the tactics used to organize the Capitol attack. Other avenues of research involve uncovering more complex coordination tactics and considering a longer time window to study how the conversations fully converge after the Capitol attack.

\subsection{Limitations}
Our work presents a few limitations. First, the rehydration allowed us to retrieve only 51.72\% of the complete dataset collected by \citet{chen_2022_2020}. This is mainly due to the content moderation and account banning performed by Twitter during the Election period and, in particular, after the moderation interventions against QAnon-specific content \cite{sharma2022characterizing}. For this reason, our findings represent a lower-bound estimate of coordinated inauthentic activity performed during the observation period. Nonetheless, despite these limitations, our analysis uncovered evidence of multiple manipulation techniques ranging from targeted amplification to manufactured consensus, which went likely undetected by Twitter.

Second, our analyses focus on common manipulation tactics, but novel and more sophisticated strategies might have been employed \cite{cinelli2022coordinated,tardelli2023temporal}. Similarly, we do not consider the activity of software-controlled accounts (i.e., bots) and state-sponsored operators \cite{luceri2020detecting,ezzeddine2022characterizing}, which might have played a role in the amplification of fringe content \cite{ferrara_2020_characterizing}. 
Finally, a considerable amount of fringe actors have migrated from Twitter to low-moderated spaces, such as alt-tech platforms like Gab and Parler, during the observation period, thus potentially affecting our results that are solely based on Twitter. This calls for broader research encompassing more diverse online platforms in the information ecosystem, which will be the objective of our next endeavors.

\subsection{Broader perspective, ethics, and competing interests}
We believe our work will have a positive broader impact on the understanding of the complex interplay between conspiracy theories and fringe narratives. These online phenomena showed potential for developing into serious offline issues, as in the case of the Capitol attack. To this end, our work can inform future efforts for curbing these and other issues.
At the same time, however, among the conceivable ethical risks of our work is the possibility that some of our findings could be used for political reasons. Additional risks may derive from the unintended disclosure of sensitive user information contained in the datasets used in this work since such information could be exploited to carry out targeted attacks. To mitigate these risks, we relied on publicly available and pseudonymized data, and we carried out analyses at an aggregated level. Moreover, the observational and retrospective nature of our analyses implies that no users were harmed as part of this research. The authors declare no competing interests. In view of scientific transparency, we make our code available on GitHub\footnote{Reference anonymized for double-blind review.} to allow reproducibility and replicability of our work. 




\bibliography{mybib}


\section{Appendix: Narratives and Hashtags}
\label{app1}

\paragraph{QAnon.} QAnon conspiracy theories involve a fringe movement that believes in an anonymous government insider and alleges a conspiracy involving Satanic child sex trafficking targeting Donald Trump. This conspiracy theory gained significant attention beyond online communities and managed to connect with major political figures. The movement's most famous slogan, ``Where we go one, we go all," gave rise to the hashtag \textbf{\#wwg1wga}. Other well-known hashtags associated with QAnon include \textbf{\#pizzagate}, which relates to the unfounded claims of child abuse by prominent Democrats, \textbf{\#taketheoath}, which encouraged QAnon followers to pledge their allegiance as ``digital soldiers" endorsed by influential members, \textbf{\#projectveritas}, referring to a far-right activist group known for engaging in disinformation campaigns, \textbf{\#thegreatawakening}, drawing parallels between the supposed battle between Trump and a cabal of satanic cannibalistic pedophiles to the concept of a biblical Great Awakening, and \textbf{\#civilwar}, which calls for a second American civil war and is often associated with white supremacist movements. Finally, \textbf{\#obamagate} emerged as a claim made by Trump, alluding to the Watergate scandal, where he alleged that former President Barack Obama wiretapped him.

\paragraph{US Election.} 
Narratives surrounding a ``stolen" election began circulating even prior to the actual vote. One notable example is the emergence of the \textbf{\#sharpiegate} narrative, which quickly spread across Twitter and various online communities. This narrative alleged that a strategy was employed in conservative precincts, where felt-tip pens (sharpies) were used to render ballots unreadable. Despite repeated debunking of the \#sharpiegate claims, the rumors gained significant traction online and even prompted Trump supporters to protest outside the Maricopa County Elections Department. The protesters chanted the slogan ``Stop the steal!" during the demonstrations, which echoed through various platforms with the hashtag \textbf{\#stopthesteal}. This hashtag eventually gave rise to a movement centered around the concept of election theft. Alongside \#stopthesteal, several other conspiracy theories emerged, each with their corresponding hashtags. The hashtags \textbf{\#hammer} and \textbf{\#scorecard} propagated claims that a software called Hammer was capable of breaching secure networks, while Scorecard allegedly manipulated the overall vote count against Trump. Similarly, the hashtag \textbf{\#hammerandscorecard} combined these two theories. The hashtag \textbf{\#dominion} became associated with the belief that Dominion Voting Systems, a manufacturer of voting machines, had ``deleted" millions of votes cast in favor of Trump. Lastly, the hoax hashtag \textbf{\#qsnatch} circulated claims that malware was actively ``stealing the election" across multiple states. 

\paragraph{COVID-19.} The hashtag \textbf{\#plandemic} is associated with a COVID-19 conspiracy theory that claims the global pandemic is a deliberately orchestrated hoax by the pharmaceutical industry and philanthropist Bill Gates. This hashtag has served as a vehicle for various anti-vaccine conspiracy theories as well. It is crucial to note that the dissemination of conspiracy theories related to the pandemic began with changes made to the electoral process in response to the COVID-19 outbreak, aiming to mitigate transmission risks.






\end{document}